\documentstyle[12pt,german,epsfig]{article}

\catcode`\@=11
\long\def\@makefntext#1{
\protect\noindent \hbox to 3.2pt {\hskip-.9pt  
$^{{\ninerm\@thefnmark}}$\hfil}#1\hfill}                

\def\@makefnmark{\hbox to 0pt{$^{\@thefnmark}$\hss}}  
        
\def\ps@myheadings{\let\@mkboth\@gobbletwo
\def\@oddhead{\hbox{}
\rightmark\hfil\ninerm\thepage}   
\def\@oddfoot{}\def\@evenhead{\ninerm\thepage\hfil
\leftmark\hbox{}}\def\@evenfoot{}
\def\sectionmark##1{}\def\subsectionmark##1{}}

\renewcommand{\thefootnote}{\fnsymbol{footnote}}

\newcounter{sectionc}\newcounter{subsectionc}\newcounter{subsubsectionc}
\renewcommand{\section}[1] {\vspace*{0.6cm}\addtocounter{sectionc}{1} 
\setcounter{subsectionc}{0}\setcounter{subsubsectionc}{0}\noindent 
        {\normalsize\bf\thesectionc. #1}\par\vspace*{0.4cm}}
\renewcommand{\subsection}[1] {\vspace*{0.6cm}\addtocounter{subsectionc}{1} 
        \setcounter{subsubsectionc}{0}\noindent 
        {\normalsize\it\thesectionc.\thesubsectionc. #1}\par\vspace*{0.4cm}}
\renewcommand{\subsubsection}[1]
{\vspace*{0.6cm}\addtocounter{subsubsectionc}{1}
        \noindent {\normalsize\rm\thesectionc.\thesubsectionc.\thesubsubsectionc. 
        #1}\par\vspace*{0.4cm}}

\newcounter{appendixc}
\newcounter{subappendixc}[appendixc]
\newcounter{subsubappendixc}[subappendixc]

\renewcommand{\appendix}[1] {\vspace*{0.6cm}
        \refstepcounter{appendixc}
        \setcounter{figure}{0}
        \setcounter{table}{0}
        \setcounter{equation}{0}
        \renewcommand{\thefigure}{\Alph{appendixc}.\arabic{figure}}
        \renewcommand{\thetable}{\Alph{appendixc}.\arabic{table}}
        \renewcommand{\theappendixc}{\Alph{appendixc}}
        \renewcommand{\theequation}{\Alph{appendixc}.\arabic{equation}}
        \noindent{\bf Appendix \theappendixc #1}\par\vspace*{0.4cm}}

\def\abstracts#1{{
        \centering{\begin{minipage}{12.2truecm}\footnotesize\baselineskip=12pt\noindent
        \centerline{\footnotesize ABSTRACT}\vspace*{0.3cm}
        \parindent=0pt #1
        \end{minipage}}\par}} 

\renewenvironment{thebibliography}[1]
        {\begin{list}{\arabic{enumi}.}
        {\usecounter{enumi}\setlength{\parsep}{0pt}
\setlength{\leftmargin 1.25cm}{\rightmargin 0pt}
         \setlength{\itemsep}{0pt} \settowidth
        {\labelwidth}{#1.}\sloppy}}{\end{list}}

\newcounter{itemlistc}
\newcounter{romanlistc}
\newcounter{alphlistc}
\newcounter{arabiclistc}

\newcommand{\fcaption}[1]{
        \refstepcounter{figure}
        \setbox\@tempboxa = \hbox{\footnotesize Fig.~\thefigure. #1}
        \ifdim \wd\@tempboxa > 6in
           {\begin{center}
        \parbox{6in}{\footnotesize\baselineskip=12pt Fig.~\thefigure. #1}
            \end{center}}
        \else
             {\begin{center}
             {\footnotesize Fig.~\thefigure. #1}
              \end{center}}
        \fi}

\newcommand{\tcaption}[1]{
        \refstepcounter{table}
        \setbox\@tempboxa = \hbox{\footnotesize Table~\thetable. #1}
        \ifdim \wd\@tempboxa > 6in
           {\begin{center}
        \parbox{6in}{\footnotesize\baselineskip=12pt Table~\thetable. #1}
            \end{center}}
        \else
             {\begin{center}
             {\footnotesize Table~\thetable. #1}
              \end{center}}
        \fi}

\def\@citex[#1]#2{\if@filesw\immediate\write\@auxout
        {\string\citation{#2}}\fi
\def\@citea{}\@cite{\@for\@citeb:=#2\do
        {\@citea\def\@citea{,}\@ifundefined
        {b@\@citeb}{{\bf ?}\@warning
        {Citation `\@citeb' on page \thepage \space undefined}}
        {\csname b@\@citeb\endcsname}}}{#1}}

\newif\if@cghi
\def\cite{\@cghitrue\@ifnextchar [{\@tempswatrue
        \@citex}{\@tempswafalse\@citex[]}}
\def\citelow{\@cghifalse\@ifnextchar [{\@tempswatrue
        \@citex}{\@tempswafalse\@citex[]}}
\def\@cite#1#2{{$\null^{#1}$\if@tempswa\typeout
        {IJCGA warning: optional citation argument 
        ignored: `#2'} \fi}}

 1
 1
 1

\font\ninerm=cmr9

\begin{document}
\centerline{\normalsize\bf OPEN HEAVY FLAVOR PRODUCTION}
\baselineskip=16pt
\centerline{\normalsize\bf IN DEEPLY INELASTIC \boldmath$ep$ 
SCATTERING AT HERA\footnote{To appear in Proc, of ``New Trends in HERA Physics'', Ringberg, May 1997}
}

\centerline{\footnotesize KARIN DAUM\footnote{presently at DESY, 
D-22603 Hamburg, Germany}}
\baselineskip=13pt
\centerline{\footnotesize\it Universit\"at Wuppertal, Rechenzentrum,
Gau\3stra\3e 20}
\baselineskip=12pt
\centerline{\footnotesize\it D-42097 Wuppertal, Germany}
\centerline{\footnotesize E-mail: daum@mail.desy.de}
\vspace*{0.3cm}
\centerline{\footnotesize (On behalf of the H1 and ZEUS collaborations)}

\vspace*{0.9cm}
\abstracts{Recent results on inclusive $D^0$ and $D^{*\pm}$ production in
deeply inelastic $ep$ scattering at $\sqrt{s}~=~301$~GeV are summarized.
The data have been collected by the H1 and ZEUS experiments at HERA. The 
total and the differential cross sections are discussed in the framework 
of LO and NLO QCD predictions. The data exhibit clear evidence for boson 
gluon fusion 
dominating open heavy flavor production in the 
kinematic range currently explored at HERA.
The measurements of $F_2^{c\overline c}(x,Q^2)$ at small Bjorken 
$x$ are presented. The prospects for future analyses of open charm and beauty
production including detector upgrades and anticipated high luminosities
are investigated.}
 
\normalsize\baselineskip=15pt
\setcounter{footnote}{0}
\renewcommand{\thefootnote}{\alph{footnote}}
\section{Introduction}
The study of heavy flavor production in deeply inelastic lepton nucleon 
scattering {\em(DIS)} is of importance for the understanding of the parton 
densities in the nucleon~\cite{theory,riemersma,harris,aivazis}. 
In accordance with early
measurements by the EMC collaboration \cite{emc}, neutral current charm quark 
production is expected to be dominated by the {\em boson gluon fusion (BGF)} 
process $\gamma(Z^0) g \rightarrow c\bar{c}$ at large Bjorken $x$. Recent
results on inclusive production of $D^0$ and $D^{*\pm}$ mesons
from H1 \cite{h1cc} and ZEUS \cite{zeuscc} at HERA enable the charm 
production mechanism at small $x$ to be
investigated.

The measurement of the charm contribution $F_2^{c\overline c}(x,Q^2)$ to the 
proton structure function $F_2$ is supposed to provide information on this mechanism.
Present parton density calculations include heavy flavors in the proton using 
different assumptions.
Charm is either considered as {\em flavor excitation (FE)} ,i.e. 
$\gamma(Z^0) q \rightarrow qX$, in the massless 
quark evolution 
starting with a charm density of zero at a 
scale $Q_0^2$ \cite{mrs,martin} or it is exclusively treated as BGF  
taking into account the charmed quark mass \cite{grv}.
In terms of phenomenological inclusive analyses \cite{martin,grv,smith} 
BGF is predicted to
dominate near threshold, while FE should lead to a good description 
well above threshold. Various schemes to match these two regions have been 
proposed\cite{aivazis,match}.
Both approaches lead to similar predictions for $F_2^{c\overline c}(x,Q^2)$ 
but differences in exclusive heavy quark spectra are expected.
 
If charmed hadron production is dominated by BGF, it
provides direct information on the gluonic content of the proton, 
$x_gg(x_g,Q^2)$, complementary to what is obtained from scaling violation 
analyses of $F_2$. Since the reconstruction of hadrons with heavy flavor 
content
unequivocally 
tags the parent quark, heavy quark spectra become directly accessible.
Compared to two jet studies
the analysis of charmed hadrons allows (a) to investigate the BGF process 
down to the phase space limit $\hat s= 4 m_c^2$ which 
extends the range for the determination of the gluon density 
in the proton significantly towards smaller $x_g$, and (b)
to tag the gluon in the proton almost free of 
background from FE of light quarks in the proton, since heavy 
flavor production from gluon splitting is suppressed by heavy quark 
masses.
\unitlength1cm
\begin{figure}[t] 
\begin{picture}(15.,5.5)
\put(0,-0.3){\epsfig{file=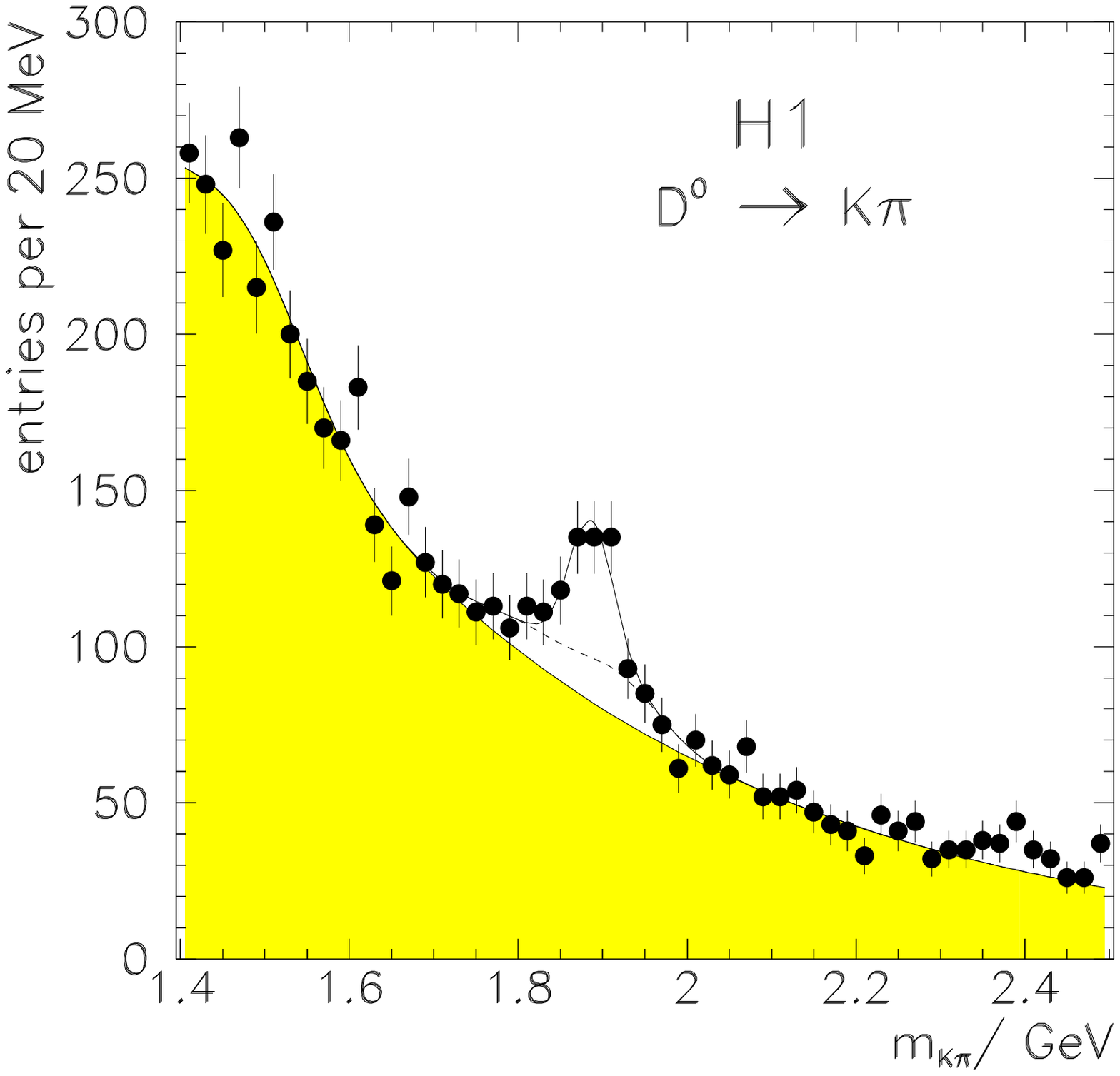,
width=6.5cm}}
\put(2.6,4.6){(a)}
\put(9.25,4.6){(b)}
\put(5.5,0.6){\includegraphics[ bb= 50 400 500 700,width=10.cm]
{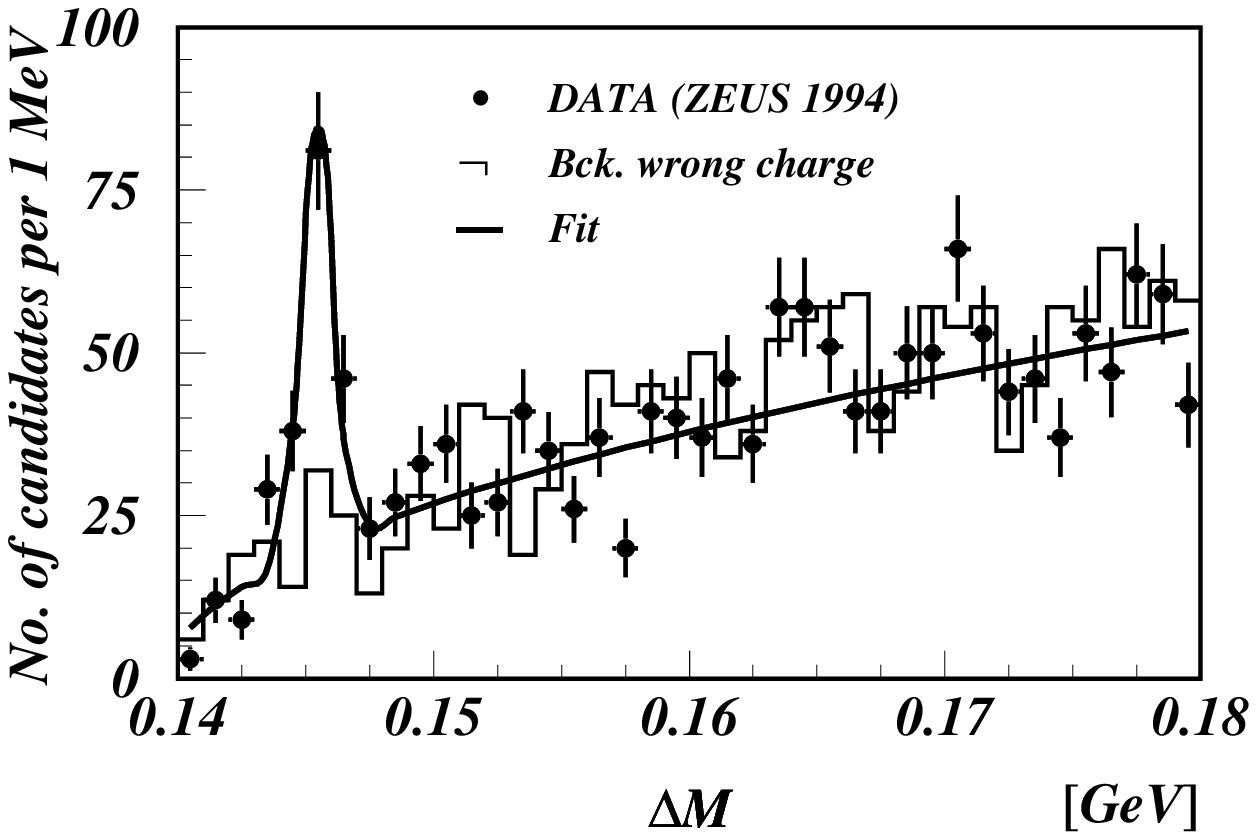}}
\end{picture}
\normalsize
\fcaption{\label{fig1}{\it(a) The $K^-\pi^+$ mass distribution observed in the
$D^0$ analysis of DIS events from H1, and (b) the $\Delta m$ 
distribution obtained in the $D^{*+}$ analysis of DIS events from ZEUS.}}
\end{figure}

\section{Results}
Published results on charm production in deeply inelastic $ep$ scattering 
exist from both experiments H1 and ZEUS based on an integrated 
luminosity of roughly 3~pb$^{-1}$ collected with each experiment
at HERA in 1994. In addition preliminary results from ZEUS are available 
based on an integrated luminosity of 6.4~pb$^{-1}$ from the HERA 1995 running. 
The H1 collaboration \cite{h1cc} has performed the tagging of heavy quark 
events by reconstructing $D^0$(1864)\footnote{Charge conjugate states are 
always included.} and $D^{*+}$(2010) mesons,
while the ZEUS collaboration \cite{zeuscc} gives results 
for the inclusive $D^{*+}$(2010) analysis.

The $D^0$ is identified via its decay mode
$D^0\rightarrow K^-\pi^+$
and the $D^{*+}$ through the decay chain
$D^{*+}\rightarrow D^0\pi^+_{slow}\rightarrow K^-\pi^+\pi^+_{slow}$.
For the latter use is made of the low $Q$-value in the decay of
$D^{*+}\rightarrow D^0\pi^+_{slow}$ 
 which leads to a better 
resolution for the mass difference
$ \Delta m = m(D^0 \pi^+_{slow}) - m(D^0)\label{deltam}$
than for the $D^{*+}$ mass itself.

Figure \ref{fig1} shows the $m_{K\pi}$ distribution obtained in the inclusive 
$D^0$ analysis of H1 and the $ \Delta m$ distribution as 
observed in the inclusive $D^{*+}$ of 
ZEUS. 
Evidently the 
number of observed events containing heavy quarks is small. Only of the order 
of 100 to 200 charmed mesons are identified in any of the different analyses.

\pagebreak
\begin{figure}[t] 
\begin{picture}(15.,5.4)
\put(5,-.3){\epsfig{file=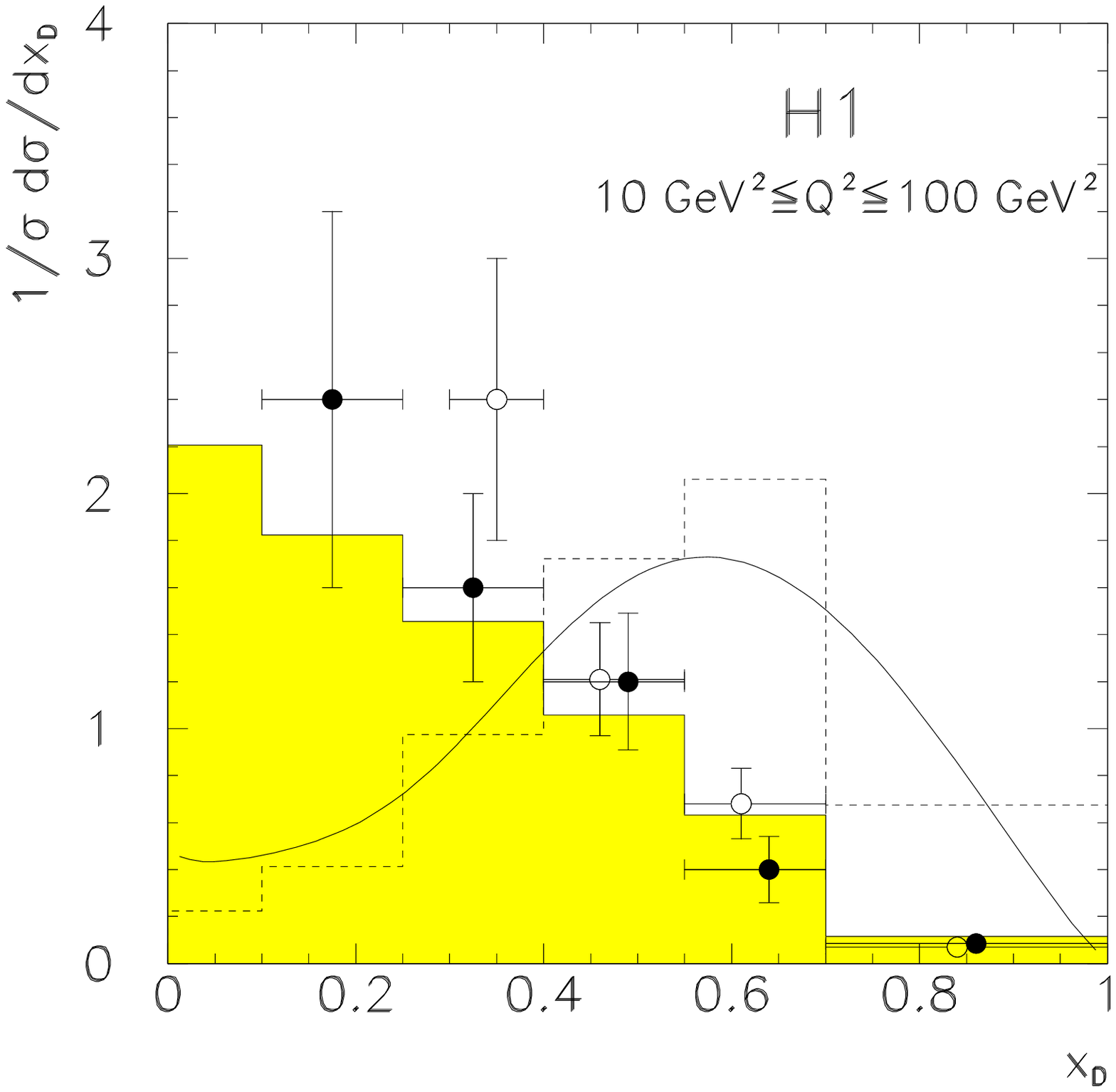,
width=6.5cm}}
\end{picture}
\normalsize
\fcaption{\label{fig2}{\it
 Normalized $x_D$ distribution in deep inelastic 
          $ep$ scattering 
          at $\langle W\rangle\approx 125$~GeV for $|\eta_D|<1.5$.
          The open/closed points represent the $D^0$/$D^*$ data 
          of H1.          
          See text for further explanations.
  }}
\end{figure}
\vspace*{-1cm}
\subsection{Mechanism of charm production in DIS}

Due to the hard fragmentation of $D$ mesons its momentum is strongly 
correlated with the parent charm quark momentum. 
Therefore the distribution of the scaled $D$ momentum in the 
$\gamma^*p$~system $x_D={2|\vec P^*_D|}/{W}$
depends on the charm quark momentum spectrum and thereby on the
mechanism of charm production.  In BGF
a $c\overline c$~pair is 
produced recoiling against the proton remnant in
the $\gamma^*p$~system, whilst in FE a single charmed quark is emitted 
opposite to the proton remnant, confining the second charmed quark 
necessary for quantum number conservation. Consequently the mean 
$\langle x_D\rangle$ of $D$ mesons in the current and central fragmentation 
region is expected to be roughly a factor of 2 smaller for BGF 
compared to FE. 

Figure~\ref{fig2} shows the distributions ${1}/{\sigma}\; {{\rm d} \sigma}/
{{\rm d}x_D}$ 
obtained from the $D^0$ and $D^{*+}$ analysis of H1 in comparison to the 
expectation for the BGF and FE process. The LO BGF prediction from AROMA
\cite{aroma}
(shaded histogram) agrees well with the data.  
The figure also includes the FE expectations for charmed mesons either by 
using the LEPTO 6.1 generator~\cite{lepto}, from which only FE 
events are selected (dashed histogram), 
or by extrapolating the results from charm production in
charged current $\stackrel{\scriptscriptstyle(-)}{\textstyle \nu}N$ scattering 
\cite{neutrino} to HERA energies (full line). 
 Large differences in the 
shape are observed for the data and these FE expectations. From a fit to the 
data of BGF and FE in arbitrary proportions it is concluded that more than 
95\% of neutral current charm production in DIS is due to the BGF process
at $\langle Q^2\rangle\approx 25$~GeV.
This observation seems to contradict recent inclusive calculations
of charm production in DIS \cite{aivazis}, from which it was concluded that for 
the kinematic range of the current analyses at HERA charmed quarks 
may already be considered as massless partons in the proton.

\begin{figure}[t]
\begin{picture}(15.,10.5)
\put(2,0){\includegraphics[ bb= 20 150 550 650,width=11.cm]
{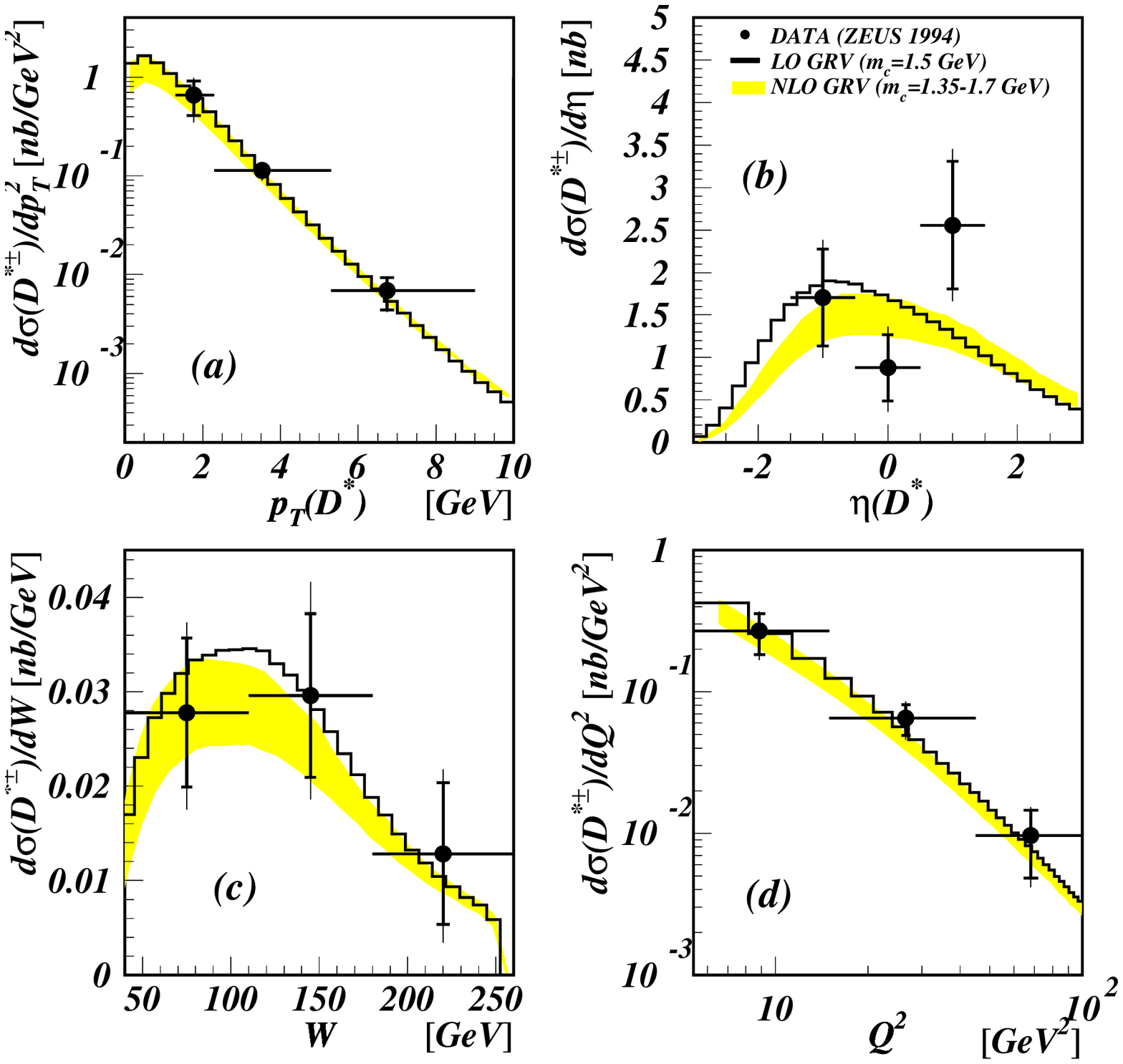}}
\end{picture}
\fcaption{\label{fig3}{\it Differential $e^+p\rightarrow D^{*\pm}X$ cross 
sections 
for $5~<~Q^2~<~100$~GeV$^2$, $y<0.7$ in the kinematic region 
$1.3~<~p_T(D^*)~<~$~GeV $|\eta(D^*)|<1.5$
as a function of a) $p_T(D^*)$, b) $\eta(D^*)$, c) $W$ and d) $Q^2$
 from ZEUS. The inner/outer error 
bars correspond to the statistical/total errors respectively.
The band and histogram are explained in the text.
}} 
\end{figure}
\subsection{ Differential inclusive $D^*$ cross sections}
Figure~\ref{fig3} shows the differential $e^+p\rightarrow D^{*\pm}X$ cross 
sections as a function of a) $p_T(D^*)$, b) $\eta(D^*)$, c) $W$ and d) $Q^2$
measured by ZEUS. The shaded band represents the result of the analytic NLO 
calculation\cite{harris} using the GRV-HO~94 \cite{grv} gluon density of 
the proton which
assumes charm being exclusively produced by BGF and treats charm quarks as 
massive particles. Both the renormalization scale and the factorization
scale are set to $\mu=\sqrt{Q^2+4m_c^2}$. The largest uncertainties in the 
calculations arise from the ignorance of $m_c$. The upper
(lower) limit corresponds to a charm quark mass $m_c=1.35(1.7)$~GeV. This
calculation reproduces well the shapes of the differential distributions. 
The figure also includes the analytic LO calculation as histograms. The LO 
calculation describes the shapes of the distributions equally well.

\subsection{Integrated charm cross sections}
All differential distributions discussed in the previous sections are found 
to be well described by the BGF process.Therefore, it is used to extrapolate 
the inclusive $D$ cross sections, $\sigma (ep\rightarrow~eDX)$, from the
visible kinematic region to the full phase space in 
order to determine the charm production cross section, 
$\sigma (ep\rightarrow~ec\overline cX)$, in DIS. 
\begin{table}[t]
\tcaption{Measured and predicted integrated charm cross sections 
$\sigma (ep\rightarrow~ec\overline cX)$ for different bins in $Q^2$ in the 
region $y<0.7$ (ZEUS) and $0.01<y<0.7$ (H1).}\label{tab1}
\small
\begin{tabular}{||l|c|r|r||}\hline\hline
{} &{} &{}&{} \\
&$m_c$~[GeV]&$5~<~Q^2~<~10$~GeV$^2$&$10~<~Q^2~<~100$~GeV$^2$\\
{} &{} &{}&{} \\
\hline
{} &{} &{}&{} \\[-3pt]
ZEUS $D^{*\pm}$ & &$13.5\pm5.2\pm1.8{+1.6\atop-1.2}$~nb &
$12.5\pm3.1\pm1.8{+1.5\atop-1.1}$~nb 
\\[-8pt]
{} &{} &{}&{}\\LO MC&1.5 &12.5~nb&14.2~nb\\LO GRV&1.5 &11.0~nb&12.4~nb\\
NLO GRV &1.5&9.4~nb&11.1~nb\\
{} &{} &{}&{}\\[-3pt]
\hline\hline
{} &{} &{}&{}\\[-3pt]
H1 $D^{*\pm}$&{}& &$15.1 \pm 1.8 {+2.4\atop-2.0}\pm1.2$~nb\\ [+3pt]
H1 $D^0$&{}&&$19.5 \pm 2.6 {+2.7\atop-2.5}{+1.6\atop-1.2}$~nb\\[-8pt]
{} &{} &{}&{} \\
NLO MRSH &1.3&&11.3~nb\\
NLO MRSH &1.7&&8.1~nb\\
H1 $F_2$ fit&1.5&&13.6~nb\\
{} &{} &{}&{} \\[-3pt]
\hline\hline
\end{tabular}
\end{table}

Table~\ref{tab1} summarizes the experimental results together with the 
predictions of the NLO calculations \cite{riemersma,harris} for 
different parton densities and different values of $m_c$. The extrapolation in 
phase space is calculated in NLO using the GRV-H0 gluon density
in case of the ZEUS 
data while the acceptance for the H1 data is calculated using the MRSH gluon 
density. In both cases $m_c=1.5$~GeV and $\mu=\sqrt{Q^2+4m_c^2}$ are chosen. 
The charm fragmentation is described by the Peterson function with
$\epsilon_c=0.035\pm0.009$ (ZEUS) and the symmetric Lund function with the 
standard setting of JETSET~7.4 (H1). The errors correspond to the statistical
error, the experimental systematic uncertainty and the model dependent 
uncertainty respectively. The latter accounts for the possible variation of the
relevant parameters of the model used for extrapolation, i.e. $xg(x,Q^2)$,
$m_c$, $\mu$ and the fragmentation function. 
In the region of overlap the cross section 
inferred from the $D^{*\pm}$ analyses are in reasonable agreement. The cross 
section from the $D^0$ analysis is higher but the few
statistics do not allow any conclusion to be drawn. 

In table~\ref{tab1} the data is also compared with predictions from analytic 
NLO calculations for different gluon densities and different values of 
$m_c$. Also given are LO analytic calculations and the result from the LO 
Monte Carlo. In addition H1 has performed a NLO calculation using the gluon
density as derived from the analysis of scaling violations of $F_2$
\cite{h1f2}. 
Within reasonable variations of the most relevant parameter, $m_c$, the NLO
calculations tend to be below the experimental results except for the gluon
density obtained from the NLO $F_2$ fit. The LO analytic calculation and the
LO MC yield cross sections generally in better agreement with the data.
\pagebreak
\begin{figure}[t]
\begin{picture}(15.,8.5)
\put(8.1,7.8){\large{\bf b)}}
\put(.3,7.8){\large{\bf a)}}
\normalsize
\put(0,-.2){\includegraphics[ bb= 0 180 600 620,width=10.5cm]
{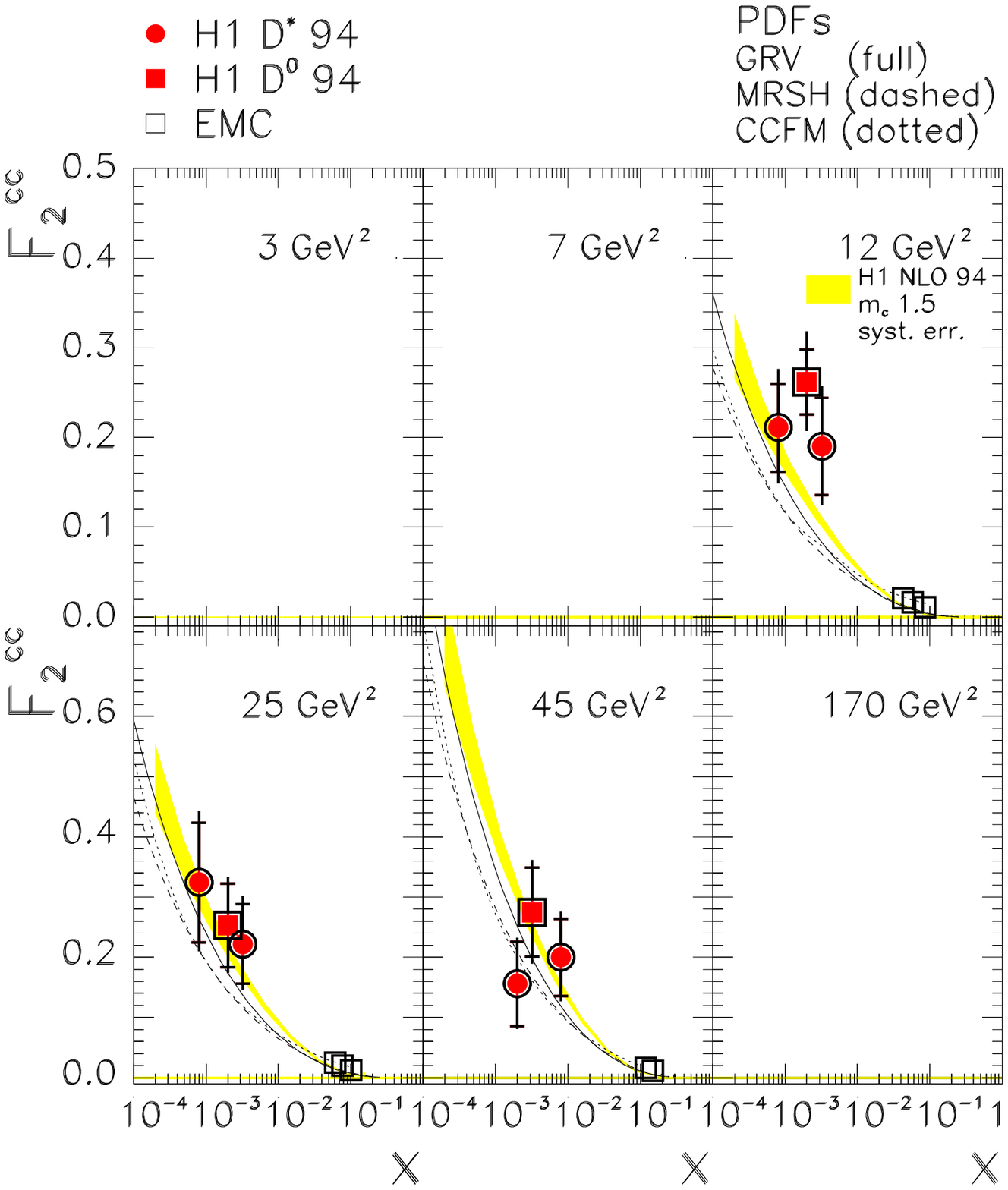}}
\put(7.8,-.2){\includegraphics[ bb= 0 180 600 620,width=10.5cm]
{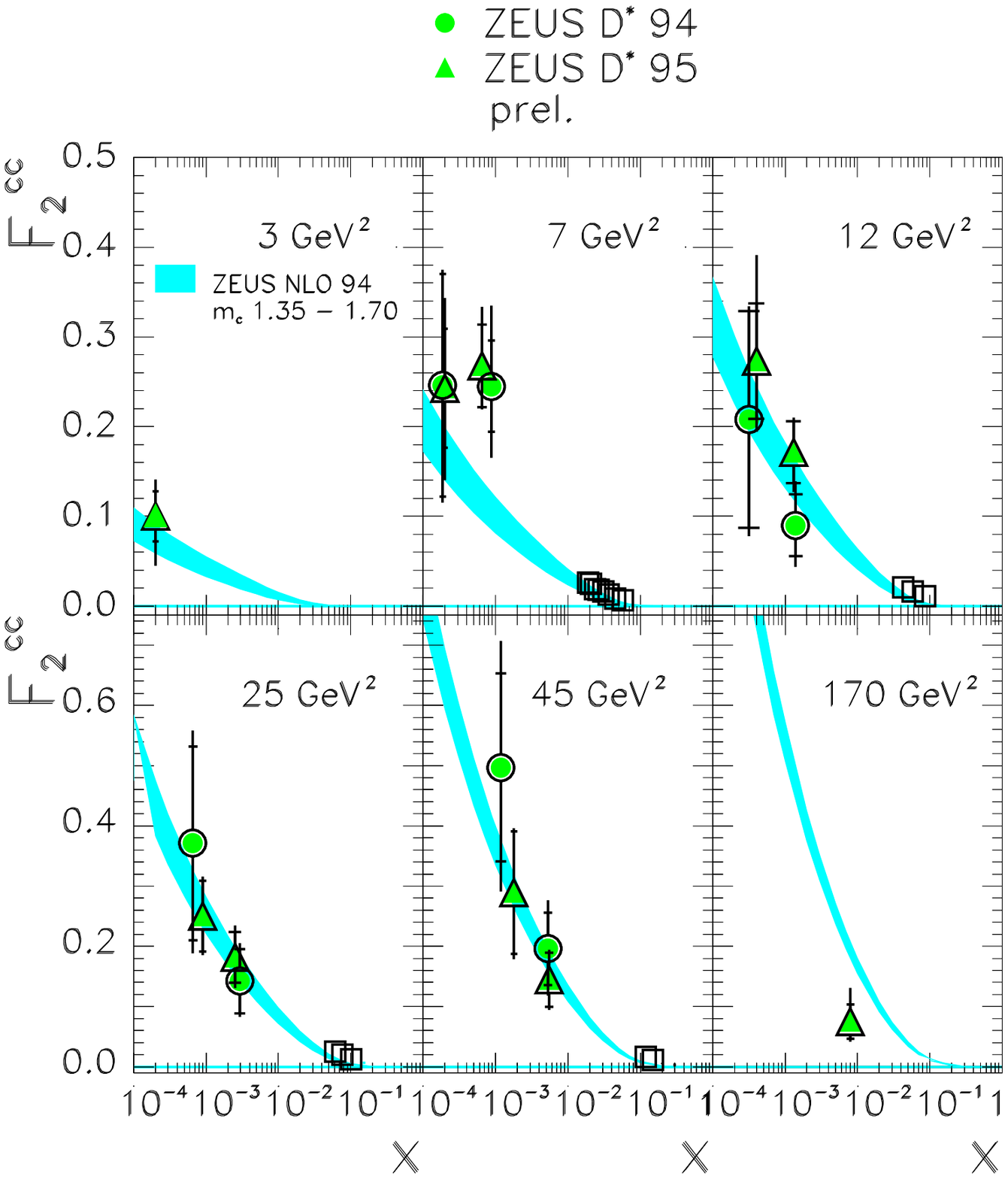}}
\end{picture}
\vspace{0.3cm}
\fcaption{\label{fig4}{\it The charm contribution $F_2^{c\overline c}$ 
          to the proton structure function as derived (a) from the 
          $D^{*+}$ (dots) and $D^0$ analysis (squares) by the 
          H1 collaboration and (b) from the $D^{*+}$ by the
          ZEUS collaboration from the 1994 data (dots) and the 1995 data
          (triangles).  The inner/outer
          error bars refer to the statistical/total errors. 
          The EMC data are also shown (open boxes).
          The shaded band represent the predictions from the NLO fits to 
          the H1 and ZEUS $F_2$ measurements respectively.
          The ZEUS 1995 data and the prediction of the NLO fit to the
          $F_2$ measurement are preliminary. 
          In (a) predictions from NLO calculations based on GRV-HO (full line),
          MRSH (dashed line) and CCFM (dotted line) gluon densities
          using a charmed quark mass of $m_c=1.5$~GeV are included 
}} 
\end{figure}
\vspace*{-1.cm}

\subsection{Charm contribution $F_2^{c\overline c}$ to the proton structure}
The charm contribution $F_2^{c\overline c}$ to the structure function is 
obtained by using the expression for the one photon exchange cross section 
for charm production
\begin{equation}
\displaystyle
\frac{{\rm d}^2\sigma^{c\overline c}}{{\rm d}x{\rm d}Q^2}=
\frac{2\pi\alpha^2}{Q^4x}
\left[1+\left(1-y\right)^2\right]\;F^{c\overline c}_2(x,Q^2)\;.
\end{equation} 
The contribution from $Z$ exchange is expected to be small in the kinematic
range explored presently. The $F_L^{c\overline c}$ contribution has been 
estimated to be negligible \cite{fl,me}. Therefore both $F_3^{c\overline c}$ 
and $F_L^{c\overline c}$ are neglected.

The $F_2^{c\overline c}$ measurements from H1 and ZEUS are displayed in 
figure \ref{fig4} together with the results from the EMC collaboration
at large $x$. Within the limited statistics the results on  
$F_2^{c\overline c}$ from H1 and ZEUS are in good agreement.
The measurements at HERA extend the range of the 
$F_2^{c\overline c}$ measurements by two orders of magnitude towards 
smaller $x$ values. The comparison of the HERA and EMC data reveals
a steep rise of $F_2^{c\overline c}$ with decreasing $x$.
H1 has derived a ratio of $F_2^{c\overline c}/F_2$ of 
\begin{equation}
\displaystyle
\left\langle F_2^{c\overline c}/F_2\right\rangle=
0.237\pm0.021\pm0.041\;, 
\end{equation}
which is an increase of about one order of magnitude
of the overall charm contribution compared to the EMC result. This is
consistent with the measured rise of the gluon distribution towards low $x$ 
and the dominance of the BGF process observed here.  
Within the accuracy of the charm data this ratio seems to be constant in 
the kinematic range explored in the H1 analysis. 

The data are also compared to the NLO calculations of $F_2^{c\overline c}$
based on the gluon densities
extracted from the NLO fits to the inclusive $F_2$. 
In figure \ref{fig4}a the prediction from
the H1 $F_2$ data,
using a charmed quark mass of $m_c=1.5$~GeV, is shown by an error band 
which includes the propagation of the statistical and the uncorrelated 
systematic errors on the total $F_2$ data through the fitting procedure.
The uncertainty in the prediction of $F_2^{c\overline c}$
due to variations of $m_c$ is indicated by the error band in figure \ref{fig4}b
for the ZEUS $F_2$ data. The upper and lower limits correspond to 
$m_c=1.35$~GeV and $m_c=1.7$~GeV respectively. No error propagation in the 
fit is performed. In the region of overlap in $Q^2$
the H1 result of this fit favors a slightly larger gluon density at small $x$ 
compared to the ZEUS measurement.

Figure \ref{fig4}a also includes the predictions from the NLO calculation 
based on GRV-HO, MRSH and CCFM \cite{ccfm} gluon density calculations. 
Although the procedures for treating heavy flavors and for determination of the
the gluonic content in the proton are quite different for these three parton 
density calculations, the resulting predictions for $F_2^{c\overline c}$ are 
very similar. The accuracy of the data in terms of statistical and systematic
uncertainty is by far not sufficient to be able to get more insight on the 
gluon in the proton from purely inclusive charm measurements.

\section{Future}
Although the charm production cross section is found to be about 25\% of the
total DIS cross section at HERA, the number of identified charm events is
fairly small at present. 
For the results presented here heavy flavor tagging is performed by the 
reconstruction of charmed hadrons. From the experimental point of view one 
advantage of this method is the small contribution to the systematic 
uncertainties due to hadronization effects.
Furthermore, as a consequence of the hard fragmentation of charm quarks, 
the reconstructed charmed hadron gives direct access to the parent charm quark.
This allows the study of the production dynamic of heavy quarks almost free of 
background from other processes. The drawback of this tagging method is the 
small tagging efficiency ($\epsilon_{tag}\approx{\cal O}(10^{-3})$)
due to the small probability that a heavy quark
fragments into a specific heavy flavored hadron which subsequently
decays into the desired 
decay mode, e.g. for the chain $c\rightarrow D^{*\pm}X\rightarrow D^0\pi^\pm_sX
\rightarrow(K^\mp\pi^\pm)\pi^\pm_sX$ this probability
is only $6.7\cdot10^{-3}$. 
\begin{figure}[t]
\begin{picture}(15.,6.5)
\put(0.5,-.3){\includegraphics[ bb= 0 50 775 475,width=13cm]
{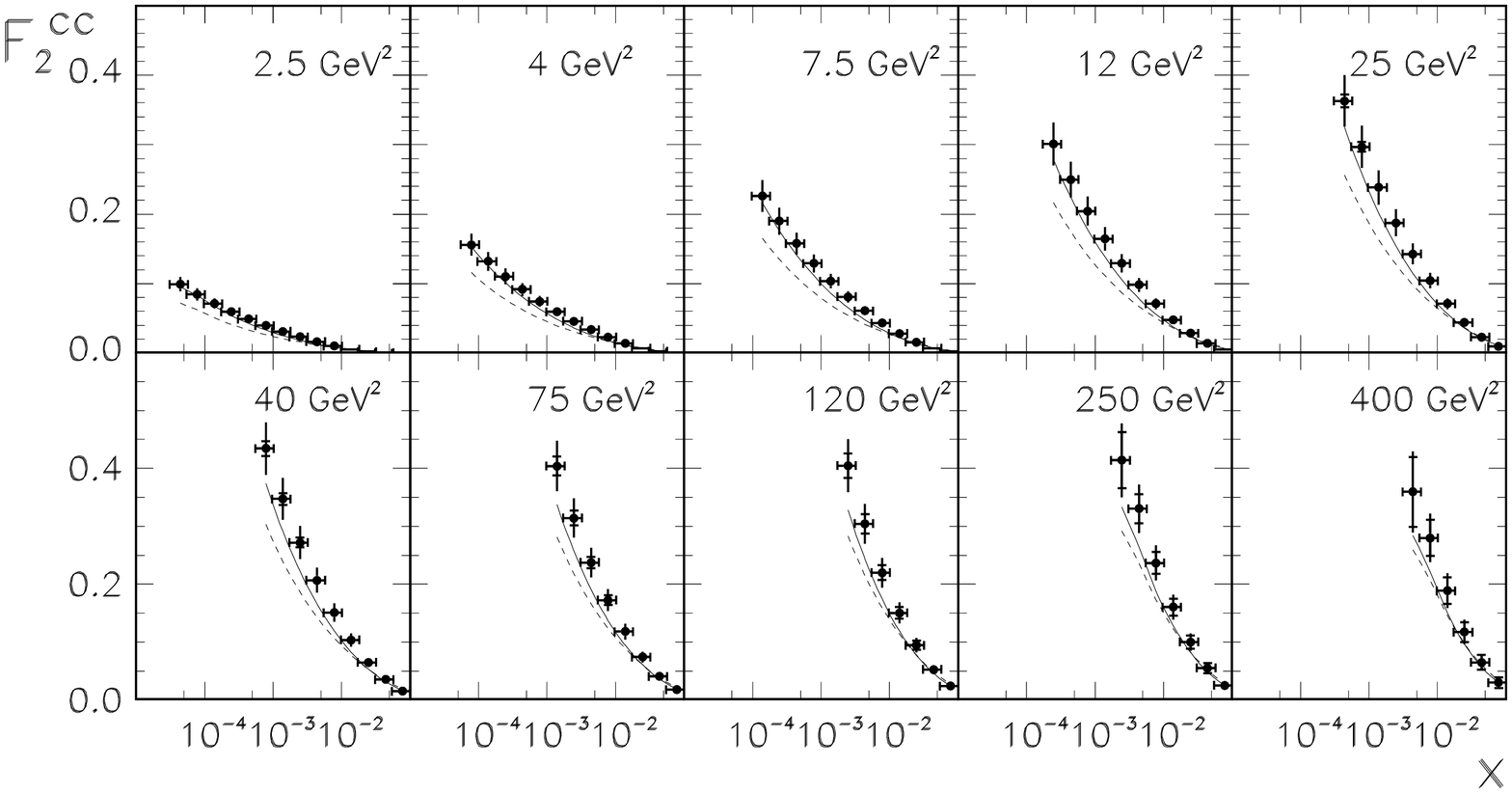}}
\end{picture}
\vspace{0.1cm}
\fcaption{\label{fig5}{\it The charm contribution $F_2^{c\overline c}$ 
as expected for an integrated luminosity of 
${\cal L}_{int}\approx 200$~pb$^{-1}$ using fully reconstructed charmed 
hadrons together with the Silicon Vertex Detector.}} 
\end{figure}

In order to increase $\epsilon_{tag}$ of heavy flavor production a Silicon 
Vertex Detector is installed in H1 this year (1997) and will be 
installed in ZEUS together with the luminosity upgrade of the HERA machine in
1999/2000. Such a device enables the reconstruction of the decay vertices of
heavy flavored hadrons and will thereby suppress the combinatorial background
in the mass reconstruction of heavy flavored hadrons.
The reduction of background will give access to a large variety of
decay modes with good signal to background ratios. Considering the analysis
of all accessible decay modes together $\epsilon_{tag}$ will be increased 
by about one order of magnitude.  

Alternatively heavy flavor events may be tagged by secondary
vertices (SVTX) having high significance without explicit reconstruction
of heavy flavored hadrons. 
This method will lead
to tagging efficiencies comparable or even superior to the efficiency expected
from the mass reconstruction method. Due to the absence of
a mass constraint the
SVTX method is expected to have larger background from faked
secondary vertices. 
A much better understanding of fragmentation, 
detector 
and reconstruction performance is needed in this case 
to achieve
experimental systematic uncertainties comparable to 
 those of the
mass reconstruction method. Furthermore it will become more difficult to access
exclusive quantities of the production dynamics since the strong correlation
between the heavy flavored hadron and the parent quark gets lost or will be
weakened in partly reconstructed heavy flavored hadrons.
\begin{figure}[t]
\begin{picture}(15.,6.5)
\put(0.5,-.4){\includegraphics[ bb= 0 50 775 475,width=13cm]
{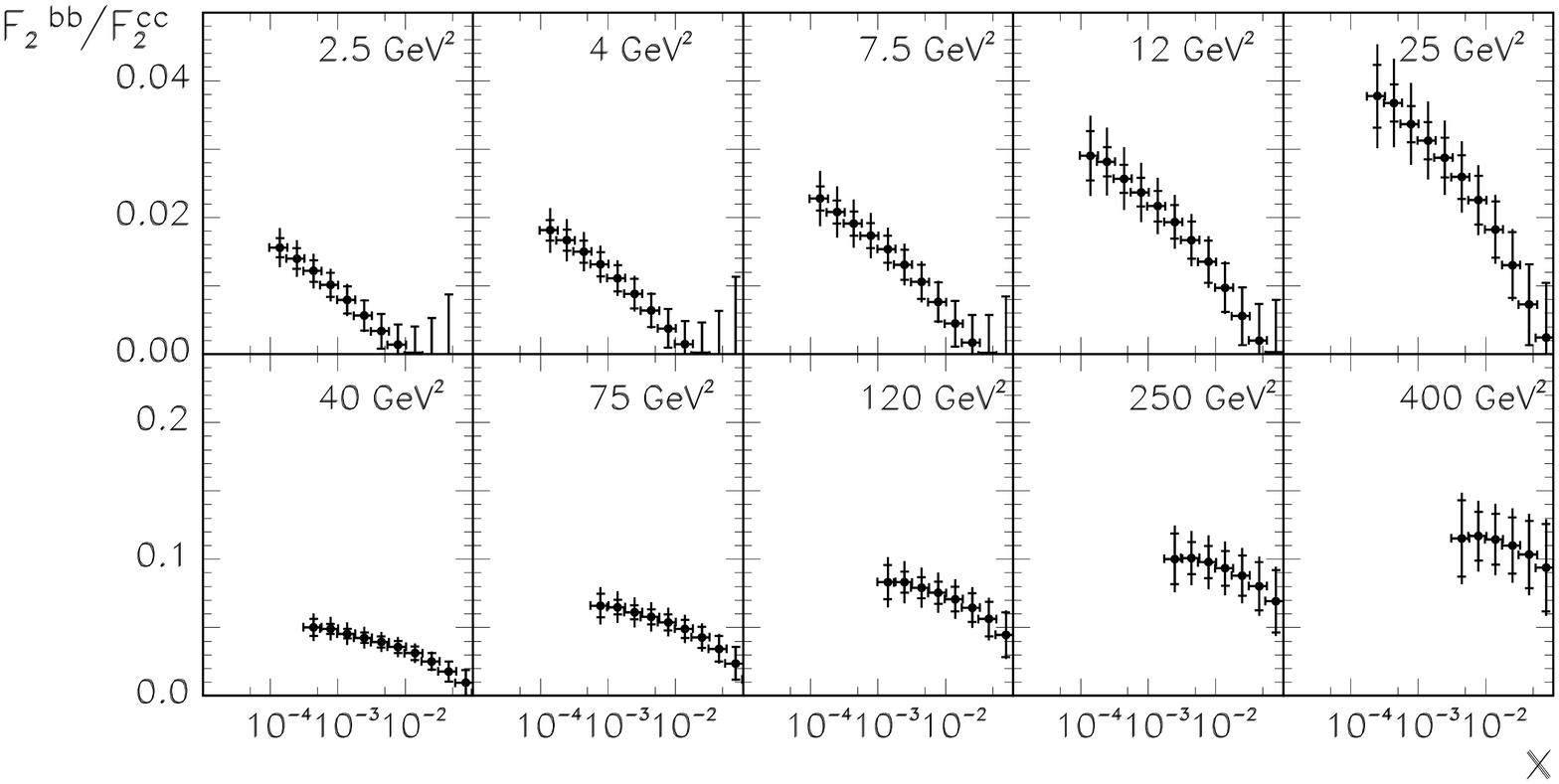}}
\end{picture}
\vspace{0.1cm}
\fcaption{\label{fig6}{\it The ratio $F_2^{b\overline b}/F_2^{c\overline c}$ 
as expected for an integrated luminosity of 
${\cal L}_{int}\approx 500$~pb$^{-1}$ using fully reconstructed charmed 
hadrons together with the Silicon Vertex Detector.}} 
\end{figure}

Figure \ref{fig5} shows the expected accuracy of $F_2^{c\overline c}$ 
measurements for an 
integrated luminosity of ${\cal L}_{int}\approx 200$~pb$^{-1}$ using fully 
reconstructed charmed hadrons together with the Silicon Vertex Detector. The 
data points 
show the expectation based on the gluon density extracted from the H1
NLO fit to the total $F_2$ for $m_c=1.5$~GeV and $\mu^2=Q^2+4m_c^2$. The 
inner/outer error bars refer to the statistical/total errors.
Also shown are the NLO predictions for the GRV-HO (full line) and 
MRSH (dashed line) parton densities and the same charm mass and scales. 
At low $Q^2$ the
measurement will be limited by the experimental uncertainty while at large 
$Q^2$ the precision will be limited by statistics. The ultimate precision of
10~\% will allow
 these parameterizations of the gluon density to be distinguished. 
 
Figure \ref{fig6} shows the expected accuracy of the ratio
 $F_2^{b\overline b}/F_2^{c\overline c}$ for an integrated luminosity of 
${\cal L}_{int}\approx 500$~pb$^{-1}$ using fully 
reconstructed charmed hadrons together with the Silicon Vertex Detector 
\cite{me}. Even for a luminosity of ${\cal L}_{int}\approx 500$~pb$^{-1}$
the measurement of this ratio will be limited by statistics in the full
$x$ and $Q^2$ plane. For a given $Q^2$ or $x$
the ratio is expected to rise with decreasing $x$ or increasing $Q^2$,
respectively. 
A mean value of
$\left\langle F_2^{b\overline b}/F_2^{c\overline c}\right\rangle
\approx 0.02$ is predicted.

\section{Conclusions}

First results on charm production in deeply inelastic $ep$ scattering at HERA
have been presented. The investigation of differential $D$ cross sections
has shown that the production dynamics of charmed mesons in the current 
and central fragmentation region may be described by the boson gluon fusion
process. The contribution from flavor excitation of charm in this region
is found to be 
negligible at $\langle Q^2\rangle\approx25$~GeV~$^2$. This observation sets
constraints on the kinematic region where charm may be treated as massless 
parton in the proton.

The measurement of 
$F_2^{c\overline c}(x,Q^2)$ 
at small $x$ reveals a strong rise of 
$F_2^{c\overline c}(x,Q^2)$ from the fixed target to the HERA regime. 
Agreement is observed between the measured $F_2^{c\overline c}$ and the results
from the NLO fits to the inclusive $F_2$. Averaged over the kinematic range
currently explored a ratio  $\left\langle F_2^{c\overline c}/F_2\right\rangle=
0.237$ is observed which is one order of magnitude larger than at large $x$.

An estimation on the accuracy for the exploration of charm and beauty
contribution  $F_2^{c\overline c}$ and $F_2^{b\overline b}$ to the structure 
of the proton for an integrated luminosity of more than 200~pb$^{-1}$
has been summarized. For $Q^2~\le~100$~GeV$^2$ the charm analysis will be 
systematically limited. The heavy flavor data will allow a direct
measurement of the gluon density in the proton to be performed. 
A precision on $xg(x,Q^2)$ 
of 10~\% may be achieved from charm data, which is competitive with other 
methods. Despite of the smallness of the ratio
$ F_2^{b\overline b}/F_2^{c\overline c}$ high luminosity running will allow
detailed studies of beauty production dynamics.

\section{References}


\begin{thebibliography}{9}
\bibitem{theory} A.\,Ali {\it et al.,} in 
{\em Proc. of the HERA Workshop}, Hamburg, Vol.\,1,393 (1988),
     A.\,Ali and D.\,Wyler, in {\it Proc. of the Workshop ``Physics at HERA''},
     Hamburg, Vol.\,2, 667 (1991) (and references therein).
\bibitem{riemersma} E. Laenen {\it et al.}, {\it Nucl. Phys.}{\bf B392}162, 
229,
S. Riemersma {\it et al.}, {\it Phys. Lett.} {\bf B347}
(1996) 143.
\bibitem{harris} B.W. Harris and J. Smith, {\it Nucl. Phys.}{\bf B452} (1995) 109, FSU-HEP-970527, [hep-ph/9706334].
\bibitem{aivazis} M.A.G. Aivazis {\it et al.}, {\it Phys. Rev.} {\bf D50} 
(1994) 3085,
3102.
\bibitem{emc} EMC Coll., J.J. Aubert {\it et al.,} 
{\it Nucl. Phys.}{\bf B231} (1983) 31.
\bibitem{h1cc} H1 Coll. C. Adloff {\it et al.}, {\it Z.
Phys.} {\bf C72} (1996) 539.
\bibitem{zeuscc} ZEUS Coll. J. Breitweg{\it  et al.},  {\it DESY preprint, 
DESY 97-089}.
\bibitem{mrs} A.D.\,Martin, R.G.\,Roberts, and W.J.\,Stirling:
        {\it Phys.Lett.} {\bf B306} (1993), 145, {\it Phys. Rev.} {\bf D50}
(1994) 6743, {\it  Phys. Rev.} {\bf D50} (1994) 6734.
\bibitem{martin} A.D.\,Martin, {\it this proceedings} (and references therein).
\bibitem{grv}
       M.\,Gl\"uck, E.\,Reya, and A.\,Vogt: {\it Z. Phys.} {\bf C 53}
 (1992) 127, M. {\it Z. Phys.} {\bf C 67} (1995) 433,
A.\,Vogt, {\it this proceedings} (and references therein).
\bibitem{smith} J. Smith, {\it this proceedings} (and references therein). 
\bibitem{match} A.D. Martin {\it et al.},DPT-96-102 and hep-ph/9612449,
 M. Buza {\it et al.}, DESY 96-258 and hep-ph/9612398.

\bibitem{aroma} G. Ingelman, J. Rathsman, and G.A. Schuler:{\it DESY preprint},
DESY 96-058, and HEP-PH/9605285. 
\bibitem{lepto} G. Ingelman: {\it``LEPTO version 6.1 - The Lund Monte Carlo for
Deep Inelastic Lepton-Nucleon Scattering''}, TSL/ISV-92-0065.
\bibitem{jetset} T. Sj\"ostrand: {\it``PYTHIA 5.7 and JETSET 7.4 Physics and
Manual''}, CERN-TH.7112/93.
\bibitem{neutrino} H. Abramowic {\it et al.}, CDHS Coll.: {\it Z. Phys.} 
{\bf C15} (1982) 19, N. Ushida {\it et al.}, E531 Coll.: {\it Phys. Lett.} 
{\bf B206} (1988) 380.
\bibitem{h1f2} S. Aid {\it et al.}, H1 Coll.:{\it  Nucl. Phys.} {\bf B470} 
(1996) 3.
\bibitem{fl} E. Leanen {\it et al.},{\it  Nucl. Phys.} {\bf B392} (1993) 162,
S. Riemersma {\it et al.}, {\it Proc. of the Workshop ``Future Physics
at HERA''}, DESY, Hamburg Vol. 1 (1995/96) 393.
\bibitem{me} K. Daum {\it et al.}, {\it Proc. of the Workshop ``Future Physics 
at HERA''}, DESY, Hamburg Vol. 1 (1995/96) 89.
\bibitem{ccfm} J. Kwieci\'nski, A.D. Martin and P.L. Sutton,  {\it Phys. Rev.}
{\bf D53} (1996) 6094, {\it Z. Phys.} {\bf C71} (1996) 585, G.P. Salam,
{\it this proceedings} (and references therein).
\end{thebibliography}
\end{document}